\documentclass[PRL,preprintnumbers,floatfix,aps,notitlepage,nofootinbib,twocolumn,showpacs,amssymb]{revtex4-1}
\usepackage{amsmath,amsfonts,bm}
\usepackage{graphicx}
\usepackage{cancel}
\usepackage[colorlinks, linkcolor={red},citecolor={blue}]{hyperref}

\begin{document}
\title{Kerr-de Sitter Black Hole Revisited}
\author{J. Ovalle}
\email[]{Corresponding author: jorge.ovalle@physics.slu.cz}
\affiliation{Research Centre for Theoretical Physics and Astrophysics,
	Institute of Physics, Silesian University in Opava, CZ-746 01 Opava, Czech Republic.}
	
\author{E. Contreras}
\email{econtreras@usfq.edu.ec}
\affiliation{Departamento de F\'isica, Colegio de Ciencias e Ingenier\'ia,
Universidad San Francisco de Quito, Quito, Ecuador.
}

\author{Z. Stuchlik}
\email[]{zdenek.stuchlik@physics.slu.cz}
\affiliation{Research Centre for Theoretical Physics and Astrophysics,
	Institute of Physics, Silesian University in Opava, CZ 746 01 Opava, Czech Republic.}

%
%
\begin{abstract}
	We interpret the cosmological constant as the energy of the vacuum, and under a minimum amount of assumptions, we show that it is deformed in the vicinity of a black hole. This leads us to reexamine the Kerr-de Sitter solution. We provide a new solution, simpler and geometrically richer, which shows the impact of the rotation in form of a warped curvature. We carry out a detailed and exact study on the new black hole solution, and we conclude with a conjecture regarding the possible impact of our results on alternative theories.
		\end{abstract} 
\maketitle
%
%
%
\section{Introduction}
In recent years, two extraordinary events have occurred, with a deep impact on theoretical physics and the way we understand the universe. These are the observational evidence that we live in an accelerated expanding Universe~\cite{Riess:1998cb,Perlmutter:1998np}, and more recently, the direct observation of a black hole (BH)~\cite{Akiyama:2019cqa}. The simplest way to explain the former is by a positive cosmological constant $\Lambda$, whereas it is well established that BHs are rotating, and therefore described by the Kerr's solution~\cite{Kerr:1963ud}. In this regard, if we want a description as realistic as possible, we should consider BHs in a de Sitter universe, i.e., $\Lambda>0$. This is precisely the scenario described by the Kerr-de Sitter (KdS) metric, a solution of Einstein field equations with cosmological constant, describing the exterior of a rotating BH~(see \textit{e.g.} Refs.~\cite{Gibbons:1977mu,Hackmann:2010zz,Akcay:2010vt,Lake:2015xca,Stuchlik:2018qyz,Li:2020drn,Stuchlik:2020rls}). This solution was discovered by Carter almost fifty years ago~\cite{Carter:1973rla}, and today we know that it represents a special case of the general Pleba\'nski-Demia\'nski metric~\cite{Plebanski:1976gy}, which is the most general solution for a Petrov Type D spacetime.

Carter builds a solution that, by definition, leaves the cosmological constant or ``vacuum energy'' immaculate, even in the strong field regime, such as the vicinity of a BH. However, the assumption of an always constant vacuum energy is a classic concept, which indeed is no longer valid in quantum field theory. It is precisely in the strong field regime where we expect that certain preestablished classical concepts begin to lose validity~\cite{Birrell:1982ix}. Of course, we can always move the cosmological constant to the matter-energy content of spacetime and postulate different models $\Lambda(x^a)$ ad hoc. This has been done for years, and it is something we definitely want to avoid. Our plan is more ambitious. We want to interpret the cosmological constant as vacuum energy, and under a minimum number of assumptions, find the way in which it is deformed in the vicinity of BHs. This deformation could give new insight to understand gravitational phenomena that cannot be explained without conjecturing exotic forms of matter.

In this paper, following the scheme explained above, we will show a gravitational effect impossible to describe by the KdS solution. This consist in a deformation undergone by the curvature $R$ of the spacetime surrounding a rotating stellar distribution. To prove what we claim, we will consider a rotating BH in an expanding universe, and show that the uniform curvature $R=-4\Lambda$ is warped around the BH~\footnote{Our signature is -2.}. To carry out the above, we depart from a spherically symmetric BH with cosmological constant, i.e., the Schwarzschild-de Sitter (SdS) solution, and then we implement the so-called gravitational decoupling (GD) approach~\cite{Ovalle:2017fgl,Ovalle:2019qyi} for axially symmetric systems~\cite{Contreras:2021yxe} to build the rotating version. Thus, we end up with a new BH solution in a de Sitter universe which is neither a $\Lambda$-vacuum solution nor does it belong to the Pleba\'nski-Demia\'nski family of metrics. The solution we found is asymptotically de Sitter, and contains the Kerr line element as a special case. In addition, it is much simpler than Carter's standard solution, and at the same time describes a richer spacetime structure, especially in the proximity of the rotating object, where we identify the aforementioned gravitational effect, i.e., the ``warped curvature''. Finally, we will see that given the simplicity of this new solution, we can carry out a detailed, analytical and exact study on the nature and structure of the horizons.
\par  
The paper is organized as follows:
in Sec.~\ref{Sgd}, we briefly review the fundamentals of the Kerr-Schild spacetimes for the spherically symmetric case, showing that for this particular symmetry the GD becomes trivial, then we generate the Schwarzschild-de Sitter solution in a simple way;
in Sec.~\ref{sec3}, we generate the axially symmetric version of the Schwarzschild-de Sitter solution, finding that this does not correspond to the known Kerr-de Sitter solution, but to a new one, whose main characteristic is the presence of a deformed curvature due to rotational effects; in Sec.~\ref{sec4}, we develop a complete and detailed analysis of BHs solutions; finally, we summarize our conclusions in Sec.~\ref{con}.

\section{Spherical Symmetry and Kerr-Schild spacetimes }
\label{Sgd}
%
%
%
Let us start from the standard Einstein field equations
\begin{equation}
	\label{efe}
	R_{\mu\nu}-\frac{1}{2}\, R\,  g_{\mu\nu}
	=
	k^2\,{T}_{\mu\nu}\ ,
\end{equation}
with $k^2=8\,\pi\,G_{\rm N}$ and $c=1$.
It is well known that the line element for all spherically symmetric and static spacetimes can be written as~\cite{Visser:1995cc}
\begin{equation}
	ds^{2}
	=
	e^{\Phi(r)}\left[1-\frac{2m(r)}{r}\right]\,dt^{2}-\frac{dr^2}{1-\frac{2m(r)}{r}}-r^2d\Omega^2
	\ ,
	\label{metric}
\end{equation}
where $\Phi(r)$ is a metric function and $m(r)$ stands for the Misner-Sharp mass function, 
which measures the amount of energy within a sphere of areal radius $r$. A particularly interesting case is that where the metric function $\Phi$ satisfies 
\begin{equation}
	\label{constr1}
\Phi(r)=0
	\ .
\end{equation}
Under the condition~\eqref{constr1}, the line element~\eqref{metric} belongs to the so-called space-times
of the Kerr-Schild class~\cite{kerrchild}, which has been extensively studied
(see \textit{e.g.} Ref.~\cite{Jacobson:2007tj}). In this case, the Einstein equations becomes 
\begin{eqnarray}
	\label{ec1ks}
	k^2\,
	T_0^{\ 0}
	&=&
	\frac{2{m}'}{r^2}\ ,
	\\
	\label{ec2ks}
	k^2\,
	T_1^{\ 1}
	&=&
	\frac{2{m}'}{r^2}\ ,
	\\
	\label{ec3ks}
	k^2\,
	T_2^{\ 2}
	&=&
	\frac{{m}''}{r}
	\ ,
\end{eqnarray}
where the energy-momentum tensor $T_{\mu\nu}$ contains an energy density ${\epsilon}
= T_0^{\ 0}$, a radial pressure ${p}_{r}=-T_1^{\ 1}$, and a tangential pressure ${p}_{t}=-T_2^{\ 2}$.
\par
There are two characteristic features of the system~\eqref{ec1ks}-\eqref{ec3ks}. The first one is the equation of state $\epsilon=-p_r$, which follows directly from Eqs.~\eqref{ec1ks} and~\eqref{ec2ks}, and the second is the linearity in (derivatives of) the mass function ${m}(r)$. Regarding the latter, we see that any solution $m(r)$ of the system~\eqref{ec1ks}-\eqref{ec3ks} can be coupled with a second one $m_s(r)$ to generate a new solution $\tilde{m}(r)$ as
\begin{eqnarray}
	\label{gdks}
	m(r)
	&\rightarrow & \tilde{m}(r) = m(r)+m_s(r)
	\ . 
\end{eqnarray}
The above represents a trivial case of the so-called gravitational decoupling~\cite{Ovalle:2017fgl,Ovalle:2019qyi}.  A simple and well-known example is the Schwarzschild-de Sitter solution, which is generated by coupling the spherically symmetric vacuum $T_{\mu\nu}=0$ with the vacuum energy of energy-momentum tensor $S_{\mu\nu}$, namely,
\begin{equation}
	\label{two}
	\tilde{T}_{\mu\nu}
	=
	\cancelto{0}{T}_{\mu\nu}+S_{\mu\nu}
	\ ,
\end{equation} 
where
\begin{equation}
	k^2\,S_{\mu\nu}=\Lambda\,g_{\mu\nu}
\end{equation} 
with $\Lambda$ the energy density of space or cosmological constant. We see, according to Eqs.~\eqref{ec1ks}-\eqref{ec3ks}, that the spherically symmetric vacuum and $\Lambda$-vacuum solution are generated, respectively, by
\begin{eqnarray}
	\label{m1}
	m&=&
	M_1\ ,
	\\
	\label{m2}
	m_s
	&=&
	M_2+\frac{\Lambda}{6}r^3
	\ ,
\end{eqnarray}
where $M_1$ and $M_2$ are constants with units of length. The two mass functions in Eqs.~\eqref{m1} and~\eqref{m2} produce, according to Eq.~\eqref{gdks}, a total mass function given by
\begin{eqnarray}
	\label{mM3}
	\tilde{m}(r)
	=
	M_1+M_2+\frac{\Lambda}{6}r^3
	\equiv
	{M}+\frac{\Lambda}{6}r^3
	\ ,
\end{eqnarray}
which plugged in Eq.~\eqref{metric}, and under the condition~\eqref{constr1}, leads to the SdS solution, namely,
\begin{equation}
	\label{SdS}
	g_{tt}=-g_{rr}=1-\frac{2{M}}{r}-\frac{\Lambda}{3}\,r^2
	\ .
\end{equation}
Notice that when $T_{\mu\nu}\neq\,0$, then $M_1\rightarrow\,\hat{m}(r)$ and the mass function $\tilde{m}$ in Eq.~\eqref{mM3} becomes
\begin{equation}
	\label{mM4}
	\tilde{m}(r)
	=
	\hat{m}(r)+M_2+\frac{\Lambda}{6}r^3
	\equiv
	m(r)+\frac{\Lambda}{6}r^3
	\ ,
\end{equation}
which plugged in Eq.~\eqref{metric} yields interior spherically symmetric space-time of the Kerr-Schild class with cosmological constant
\begin{equation}
	\label{SdS2}
	g_{tt}=-g_{rr}=1-\frac{2{m(r)}}{r}-\frac{\Lambda}{3}\,r^2
	\ .
\end{equation}
In summary, Eqs.~\eqref{SdS} and~\eqref{SdS2} represent, respectively, the exterior and interior solution for a  self-gravitating object in a spherically symmetric de Sitter vacuum.
\section{Axially symmetric case}
\label{sec3}
In order to generate the rotating version of the Schwarzschild-de Sitter line element in Eq.~\eqref{SdS}, we follow the strategy described in Ref.~\cite{Contreras:2021yxe}. Let us start with the Kerr-Schild metric in Boyer-Lindquist coordinated, namely, the Gurses-Gursey metric~\cite{Gurses:1975vu}
\begin{eqnarray}
\label{kerrex}
ds^{2}
&=&
\left[1-\frac{2\,r\,{m}(r)}{{\rho}^2}\right]
dt^{2}
+
\frac{4\, {a}\, r\,{m}(r)\, \sin^{2}\theta}{{\rho}^{2}}
\,dt\,d\phi
\nonumber
\\
&&
-
\frac{{\rho}^{2}}{{\Delta}}\,dr^{2}
-
{\rho}^{2}\,d\theta^{2}
-
\frac{{\Sigma}\, \sin^{2}\theta}{{\rho}^{2}}\,d\phi^{2}
\ ,
\end{eqnarray}
with
\begin{eqnarray}
{\rho}^2
&=&
r^2+{a}^{2}\cos^{2}\theta\ ,
\label{f0}
\\
{\Delta}
& = &
r^2-2\,r\,{m}(r)
+{a}^{2}\ ,
\label{f2}
\\
{\Sigma}
& = &
\left(r^{2}+{a}^{2}\right)^{2}
-{\Delta}\, a^2\sin^{2}\theta\ ,
\label{f3}
\end{eqnarray}
and
\begin{equation}
{a}\,=\,{J}/{M}\ ,
\end{equation}
where ${J}$ is the angular momentum and ${M}$ the total mass of the system. The line element~\eqref{kerrex} represents the simplest nontrivial extension of the Kerr metric, and it reduces to the Kerr solution when the metric function ${m}={M}$. The metric~\eqref{kerrex} is the rotational version of the spherically symmetric one in Eq.~\eqref{metric} under the constraint~\eqref{constr1}. However, we have to bear in mind that the definition of the coordinate $r$ in Eq.~\eqref{kerrex} is not the usual, but given by
\begin{equation}
	\frac{x^2+y^2}{r^2+a^2}+\frac{z^2}{r^2}=1\ .
\end{equation}
\par

A critical feature of the metric~\eqref{kerrex}, inherited from its spherical version given by Eqs.~\eqref{metric} and~\eqref{constr1}, is the linearity of the Einstein tensor in the metric function $m(r)$, which explicitly reads
\begin{eqnarray}
\label{ec1a}
{G}_0^{\ 0}
&=&
2\,\frac{r^4+\left(\rho^2-r^2\right)^2+{a}^2\left(2\,r^2-\rho^2\right)}{\rho^6}
{m}'
\nonumber
\\
&&
-\frac{r\,{a}^2\,\sin^{2}\theta }{\rho^4}
{m}'' \ ,
\\
\label{ec2a}
{G}_1^{\ 1}
&=&
2\,\frac{r^2}{\rho^4}
\,{m}'
\ ,
\\
\label{ec3a}
{G}_2^{\ 2}
&=&
2\,\frac{\rho^2-r^2}{\rho^4}
\,{m}'
+\frac{r}{\rho^2}
{m}'' \ ,
\\
\label{ec4a}
{G}_3^{\ 3}
&=&
2\,\frac{2\,r^2\left(\rho^2-r^2\right)+{a}^2\left(\rho^2-2\,r^2\right)}{\rho^6}
\,{m}'
\nonumber
\\
&&
+\frac{r\left({a}^2+r^2\right)}{\rho^4}
\,{m}''
\ ,
\\
\label{ec5a}
{G}^{\ 3}_{0}
&=&
2\,\frac{{a}\left(2\,r^2-\rho^2\right)}{\rho^6}
\,{m}'
-\frac{{a}\,r}{\rho^4}
\,{m}''
\ . 
\end{eqnarray}
Hence, as the spherically symmetric case, two different solutions $m(r)$ and $m_s(r)$ can be coupled by Eq.~\eqref{gdks} to generate a new one $\tilde{m}(r)$. However, as noticed in Ref.~\cite{Contreras:2021yxe}, the coupling~\eqref{gdks} must be complemented by the requirement
\begin{equation}
	\tilde{a}=a=a_s \ ,
\end{equation}
where $\{a,\,a_s,\,\tilde{a}\}$ are the rotational parameters associated, respectively, with the mass functions $\{m,\,m_s,\,\tilde{m}\}$. 
%
\subsection*{New Kerr-de Sitter solution}
\label{new}

Before introducing the new KdS solution, let us briefly recall the standard one discovered by Carter~\cite{Carter:1973rla}, whose line element is given by
\begin{eqnarray}
	\label{kdsstandard}
	ds^{2}
	&=&
	\left[\frac{\Delta_r-\Delta_\theta\,a^2\sin^{2}\theta}{{\rho^2\,\Xi^2}}\right]
	dt^{2}-
	\frac{{\rho}^{2}}{{\Delta_r}}\,dr^{2}-
	\frac{{\rho}^{2}}{{\Delta_\theta}}\,d\theta^{2}
	\nonumber
	\\
	&&
	-\frac{\sin^{2}\theta}{{\rho}^{2}\,\Xi^2}\left[{{\Delta_\theta\left(r^{2}+{a}^{2}\right)^{2}
			-{\Delta_r}\, a^2\sin^{2}\theta}}\right]\,d\phi^{2}
	\nonumber
	\\
	&&
	+
	\frac{2\, {a}\, \sin^{2}\theta}{{\rho}^{2}\,\Xi^2}\left[\Delta_\theta(r^2+a^2)-\Delta_r\right]
	\,dt\,d\phi\ ,
\end{eqnarray}
with
\begin{eqnarray}
	{\Delta_r}
	& = &
	r^2-2\,M\,r
	+{a}^{2}-\frac{\Lambda}{3}\,r^2(r^2+a^2)\ ,
	\label{dr}
	\\
	{\Delta_\theta}
	& = &
	1+\frac{\Lambda}{3}a^2\cos^{2}\theta\ ,
	\label{dt}
	\\
	{\Xi}
	& = &
	1+\frac{\Lambda}{3}a^2
	\label{xi}
\end{eqnarray}
This is a $\Lambda$-vacuum solution of the Einstein field equations with cosmological constant and therefore satisfies 
\begin{equation}
	\label{einst-L}
	R_{\mu\nu} = -\Lambda\,g_{\mu\nu}
	\ .
\end{equation}
A complete study of the solution~\eqref{kdsstandard} can be found in Refs.~\cite{Gibbons:1977mu,Hackmann:2010zz,Akcay:2010vt,Lake:2015xca,Stuchlik:2018qyz,Li:2020drn,Stuchlik:2020rls}.
\par
Now, we proceed to generate the new solution. We start by identifying the mass function of the spherically symmetric seed solution. This is the de Sitter-Schwarzschild mass function given by Eq.~\eqref{mM3}, which plugged into the metric~\eqref{kerrex} leads to
\begin{eqnarray}
	\label{newkerrds}
	ds^{2}
	&=&
	\left[\frac{\Delta_\Lambda-a^2\,\sin^{2}\theta}{\rho^2}\right]
	dt^{2}	-
	\frac{{\rho}^{2}}{{\Delta_\Lambda}}\,dr^{2}
	\nonumber
	\\
	&&
	-
	{\rho}^{2}\,d\theta^{2}
	-
	\frac{{\Sigma_\Lambda}\, \sin^{2}\theta}{{\rho}^{2}}\,d\phi^{2}
	\ ,
	\nonumber
	\\
	&&
	+
	\frac{2\, {a}\sin^{2}\theta}{{\rho}^{2}}\left(r^2+a^2-\Delta_\Lambda\right) 
	\,dt\,d\phi\ ,
\end{eqnarray}
with
\begin{eqnarray}
	{\Delta_\Lambda}
	& = &
	r^2-2\,M\,r
	+{a}^{2}-\frac{\Lambda}{3}\,r^4\ ,
	\label{f22}
	\\
	{\Sigma_\Lambda}
	& = &
	\left(r^{2}+{a}^{2}\right)^{2}
	-{\Delta_\Lambda}\,a^2\sin^{2} \theta\ ,
	\label{f33}
\end{eqnarray}
 and $\rho$ defined in Eq.~\eqref{f0}. The line element~\eqref{newkerrds} is a new solution of the Einstein field equations describing the exterior of a rotating stellar object in a de Sitter or anti-de Sitter background. As far as we know, this solution has never been reported before. However, it is fair to mention that in Refs.~\cite{Burinskii:2001bq,Dymnikova:2006wn,Smailagic:2010nv,Bambi:2013ufa,Azreg-Ainou:2014nra,Dymnikova:2016nlb} the same strategy was used to generate rotating regular BHs from a spherically symmetric seed solution. In this respect, notice that the interior of a rotating distribution with cosmological constant can be described by Eq.~\eqref{newkerrds} but substituting $M\rightarrow\,m(r)$. This corresponds to the use of the interior de Sitter-Schwarzschild mass function in Eq.~\eqref{mM4} [instead of Eq.~\eqref{mM3}].
 \par
We see that the metric~\eqref{newkerrds} looks quite simpler than the line element~\eqref{kdsstandard}. However, a simpler line element does not necessarily mean a less rich space-time structure, as we will see below. First of all, we can assure that both solutions are different, since the KdS metric in Eq.~\eqref{kdsstandard} is a $\Lambda$-vacuum solution, whereas the metric in Eq.~\eqref{newkerrds} is not.
In fact, we find that the curvature 
\begin{equation}
	\label{R}
	R=-4\,\Lambda\,\frac{r^2}{\rho^2}\neq-4\,\Lambda
\end{equation}
for the line element in Eq.~\eqref{newkerrds}, and therefore it is not a solution of Eq.~\eqref{einst-L}. Notice that the curvature is warped everywhere but in the equatorial plane, where remains constant. The warped effect is particularly significant near the rotating distribution, i.e., $r\sim\,a$ and disappears far enough, where $R\sim-4\Lambda$ for $r>>a$. This effect will never appear in a KdS space time, since by construction it is a constant-curvature solution. We conclude that the line element~\eqref{newkerrds} is necessary to elucidate the effects of the rotating object in its immediate surroundings.
\par
Regarding the relationship between $R$ and $\Lambda$, note that we can write the curvature in Eq.~\eqref{R} as
\begin{equation}
	\label{R2}
	R(r,\theta)=-4\,\tilde{\Lambda}(r,\theta)\ ,
\end{equation}
where we have introduced the effective cosmological constant
\begin{equation}
	\label{L-effective}
	\tilde{\Lambda}(r,\theta)\equiv\Lambda\,\frac{r^2}{\rho^2}=\Lambda\left[\frac{r^2}{r^2+{a}^{2}\cos^{2}\theta}\right]\ .
\end{equation}
The expression in Eq.~\eqref{L-effective} clearly shows the rotational effect on vacuum energy. We see that $\tilde{\Lambda}\rightarrow\,\Lambda$ for $r\,\gg\,a$. Also notice that for $r\neq\,0$
\begin{equation}
	\label{Lmaxmin}
	\Lambda\left(\frac{r^2}{r^2+a^2}\right)\,\leq\,\tilde{\Lambda}\,\leq\,{\Lambda}\ ,
\end{equation}
where $\tilde{\Lambda}_{\rm max}$ and $\tilde{\Lambda}_{\rm min}$ in Eq.~\eqref{Lmaxmin} occur in the equatorial plane and axis of rotation  respectively. We emphasize that neither $\Lambda$ nor $\tilde{\Lambda}$ is the energy of the system. The reason is that the metric in Eq.~\eqref{newkerrds} is not a solution of Eq.~\eqref{einst-L} but Einstein field equations~\eqref{efe}, where ${T}_{\mu\nu}$ generating the metric~(\ref{newkerrds})
is given by
\begin{eqnarray}\label{tmunu}
	{T}^{\mu\nu}
	=
	{\epsilon}\, {u}^{\mu}\,{u}^{\nu}
	+{p}_{r}\,{l}^{\mu}\,{l}^{\nu}
	+{p}_{\theta}\,{n}^{\mu}\,{n}^{\nu}
	+{p}_{\phi}\,{m}^{\mu}\,{m}^{\nu}
	\ ,
\end{eqnarray}
where the orthonormal tetrad $\{{u}^{\mu},\,{l}^{\mu},\,{n}^{\mu},\,{m}^{\mu}\}$ reads~\cite{Gurses:1975vu}
\begin{eqnarray}
	{u}^{\mu}
	&=&
	\frac{(r^{2}+{a}^{2})\delta^\mu_0+a\,\delta^\mu_3}{\sqrt{\rho^{2}\Delta}}
	\ ,
	\qquad
	{l}^{\mu}
	=
	\sqrt{\frac{\Delta}{\rho^{2}}}\,\delta^\mu_1
	\label{2}
	\nonumber
	\\
	{n}^{\mu}
	&=&
	\frac{1}{\sqrt{\rho^{2}}}\,\delta^\mu_2
	\ ,
	\qquad
	{m}^{\mu}
	=
	-\frac{{a}\sin^{2}\theta\,\delta^\mu_0+\delta^\mu_3}{\sqrt{\rho^{2}}\sin\theta}
	\ ,
	\label{4}
\end{eqnarray}
and the energy density ${\epsilon}$ and pressures ${p}_r$, ${p}_\theta$, and ${p}_\phi$ satisfy
\begin{eqnarray}
	\label{energyax}
	{\epsilon}
	&=&
	-{p}_{r}
	=
	\frac{2\,r^2}{\rho ^4}\, {m}'=
	\Lambda\,\frac{r^4}{\rho^4}=\frac{\tilde{\Lambda}^2}{\Lambda}\ ,
	\\
	\label{pressuresax}
	{p}_{\theta}
	&=&
	{p}_{\phi}
	=
	-\frac{r }{\rho ^2}\,{m}''
	+\frac{2\left(r^2-\rho^2\right)}{\rho ^4} \,{m}'\nonumber\\
	&=&\epsilon-2\,\Lambda\,\frac{r^2}{\rho^2}
	\ .
\end{eqnarray}
%
It is quite easy to check, from Eqs.~\eqref{energyax} and~\eqref{pressuresax}, that the dominant energy condition
\begin{eqnarray}
	{\epsilon}
	&\geq&
	0\ ,
	\label{dom1}
	\\
	\epsilon
	&\geq&
	|{p}_{i}|
	\quad
	\left(i=r,\theta,\phi\right)
\end{eqnarray}
holds for $\Lambda\,>\,0$, but is violated for $\Lambda\,<0$, whereas the strong energy condition
\begin{eqnarray}
	\nonumber
	&&
	{\epsilon}+{p}_{r}+2\,{p}_{\theta}
	\geq
	0
	\\
	\label{strong01}
	&&
	{\epsilon}+{p}_{r}
	\geq
	0
	\\
	&&
	{\epsilon}+{p}_{\theta}
	\geq
	0
	\nonumber
\end{eqnarray}
is satisfied for $\Lambda\,<\,0$ but violated for $\Lambda\,>0$.
\section{Black holes}
\label{sec4}
Notice that the metric~\eqref{newkerrds} becomes singular if $\rho=0$ or $\Delta_\Lambda=0$. The first case is the ring singularity of the Kerr solution, and represents a physical (curvature) singularity\footnote{We see from Eq.~\eqref{R} that the curvature $R$ is regular for $r\rightarrow\,0$. However, the Kretschmann scalar is singular at $(r=0,\,\theta=\pi/2)$.}. However, this singularity can be removed when $M\rightarrow\,m(r)$. The second case is a coordinate singularity which indicates a horizon. 

The equation determining the horizon of the metric~\eqref{newkerrds}
is given by $0=\tilde g^{rr}\sim\Delta_\Lambda$, which yields
\begin{equation}
	\label{hkds}
	r^4-\frac{3}{\Lambda}r^2+\frac{6\,M}{\Lambda}r-\frac{3\,a^2}{\Lambda}
	=
	0
	\ .
\end{equation}
We see that the Kerr horizon 
\begin{equation}
r_{\rm Kerr}={ M}+\sqrt{{ M}^2-a^2}
\end{equation}
 is recovered for $\Lambda=0$, and the cosmological horizon of the de Sitter solution, with characteristic length $\ell=\sqrt{3/\Lambda}$, is recovered for $a=0$ and $r>>M$. 
 
 The roots of Eq.~\eqref{hkds} can be expressed by 
\begin{equation}
	\label{formal}
	(r-r_+)(r-r_-)(r-r_{++})(r-r_{- -})=0\ ,
\end{equation}
where $r_{++}\,>\,r_+>\,r_-\,>\,r_{- -}$ are, respectively, the cosmological horizon, the event horizon, the Cauchy horizon, and inner cosmological
horizon. From Eq.~\eqref{f22}, we see that the quartic equation $\Delta_\Lambda=0$ contains the free parameters $\{M,\,a^2,\,\Lambda\}$ of the solution~\eqref{newkerrds}, and therefore $r_i=r_i(M,\,a^2,\,\Lambda)$ for each of the horizons in Eq.~\eqref{formal}. They are relatively simple, and explicitly given by
\begin{figure}[h!]
	\centering
	\includegraphics[scale=0.24]{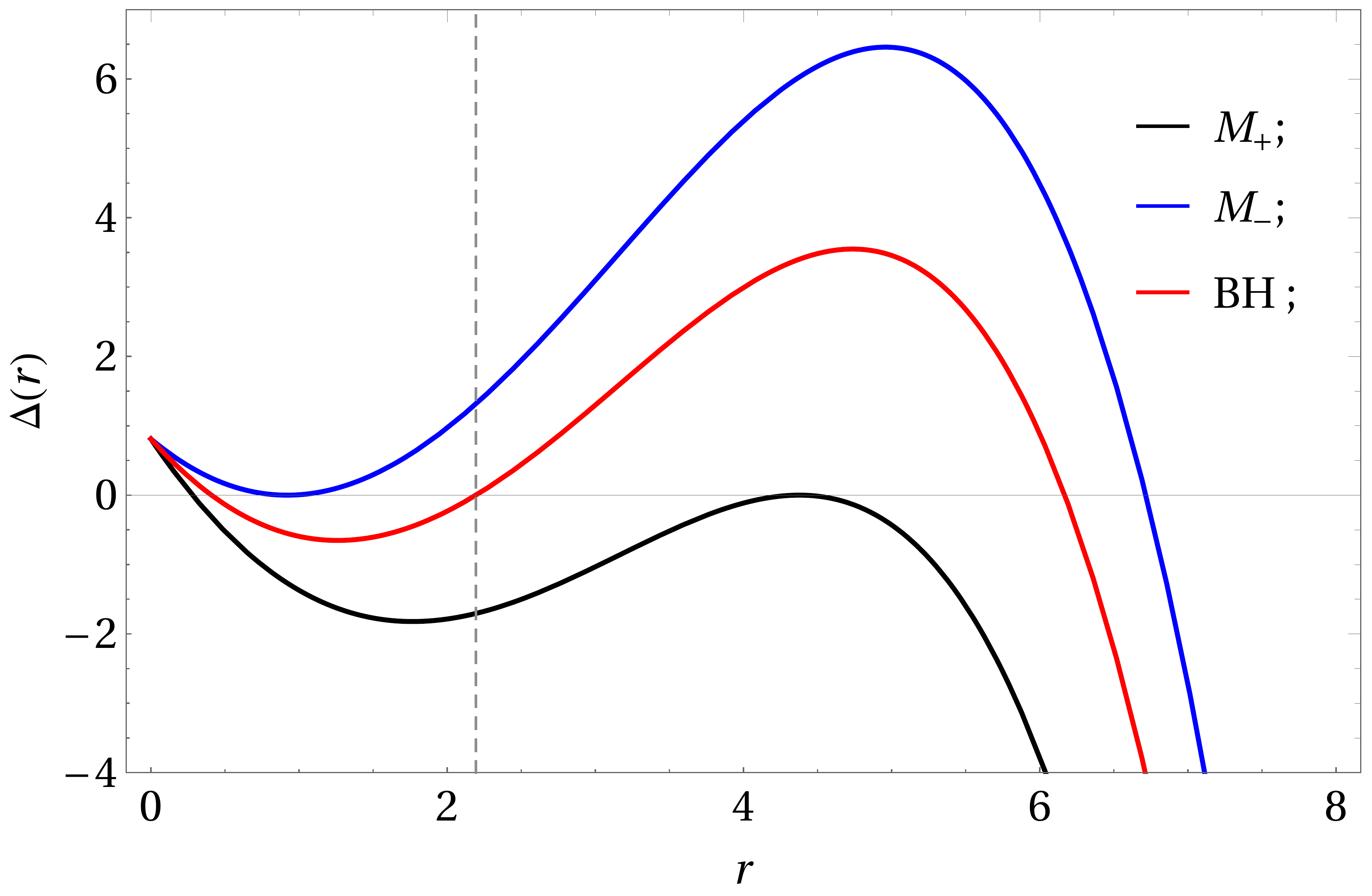}
	\caption{\label{roots} Metric function $\Delta$ for the two bounds shown in Eq.~\eqref{bounds}, i.e., extremal BH ($M_-$) and maximum BH ($M_+$), and for a BH with mass $M$ between the two bounds. We take $\{a,\,\Lambda\}=\{0.9,\,0.05\}.$ The vertical dashed line shows the event horizon $r=r_+$.}
	\label{fig1}
\end{figure}
\begin{eqnarray}
	\label{roots}
	&&r_\pm=\pm\frac{1}{2}\sqrt{\gamma}\mp\frac{1}{2}\sqrt{\frac{6}{\Lambda}-\gamma\mp\frac{12M}{\Lambda\sqrt{\gamma}}}\ ,\nonumber\\
	&&r_{++}=\frac{1}{2}\sqrt{\gamma}+\frac{1}{2}\sqrt{\frac{6}{\Lambda}-\gamma-\frac{12M}{\Lambda\sqrt{\gamma}}}\ ,
	\nonumber\\
	&&r_{--}=-\frac{1}{2}\sqrt{\gamma}-\frac{1}{2}\sqrt{\frac{6}{\Lambda}-\gamma+\frac{12M}{\Lambda\sqrt{\gamma}}}\ ,
\end{eqnarray}
with
\begin{eqnarray}
	&&\gamma=\frac{2}{\Lambda}+\frac{\alpha}{\Lambda\beta}+\frac{\beta}{\Lambda}\ ,\,\,\,\alpha=1-4a^2\Lambda\ ,\,\,\,\,\beta=(Y-X)^{1/3}\ ,\nonumber\\
	&&X=1+12a^2\Lambda-18M^2\Lambda\ ,\nonumber\\
	&&Y=2\sqrt{\Lambda[a^2(3+4a^2\Lambda)^2+9M^2(9M^2\Lambda-12a^2\Lambda-1)]}\ .\nonumber
\end{eqnarray}
For the four roots in Eq.~\eqref{roots} to be real, $\Lambda>0$ and $D>0$, with $D$ being the discriminant of the polynomial equation~\eqref{hkds}, which can be written as
\begin{equation}
	D=-\frac{108}{\Lambda^6}Y^2\ .
\end{equation}
The condition $D>0$ is satisfied for a specific range of values for $M^2=M^2(\Lambda,a^2)$. This leads to an upper ($M_{\rm max}\equiv\,M_{+}$) and lower ($M_{\rm min}\equiv\,M_{-}$) bound for $M$, given by
\begin{equation}
	\label{bounds}
	M^2_{\pm}=\frac{1+12a^2\Lambda\pm(1-4a^2\Lambda)^{3/2}}{18\Lambda}\ ,
\end{equation}
which are found by solving the degenerated case $D=0$. This occurs, respectively, at
 \begin{equation}
 	\label{bounds2}
 h_{\rm D_{\pm}}=\frac{1}{\sqrt{2\Lambda}}\sqrt{1\pm\sqrt{1-4a^2\Lambda}}\ .
 \end{equation}
We want to emphasize that, contrary to the KdS solution, the expressions in Eqs.~\eqref{bounds} and~\eqref{bounds2} are simple and exact. There are not black holes beyond these bonds. Indeed, solutions with $M<M_{\rm min}$ or $M>M_{\rm max}$ describe, respectively, a ring singularity enclosed between two cosmological horizons, or between a inner cosmological horizon and a Cauchy horizon, as we can see in Fig.~\ref{fig1}.
\begin{figure}[h!]
	\includegraphics[scale=0.18]{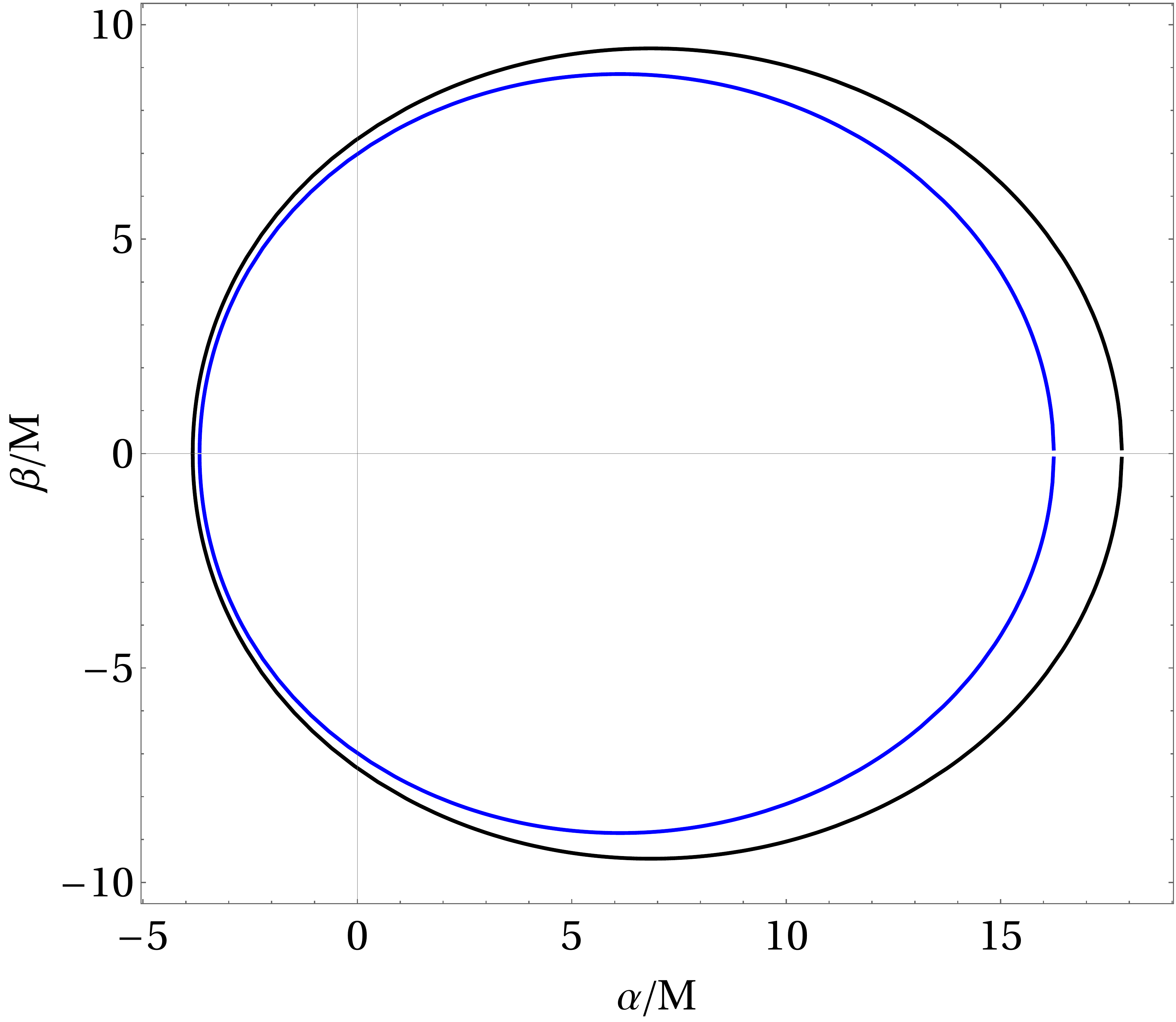}
	\caption{Shadows for the KdS solution~\eqref{kdsstandard} (black) and for the new solution~\eqref{newkerrds} (blue),  for $M=1.06,\,a=0.9,\,\Lambda=0.06$. The parameters $\alpha,\,\beta$ are the celestial coordinates~\cite{Vazquez:2003zm}.}
	\label{fig2}
\end{figure}

 For the case $M_{\rm min}<M<M_{\rm max}$, we have the event horizon $r_+$, two cosmological horizons $r_{--}$ and $r_{++}$, as well as the Cauchy horizon $r_-$. For $M=M_{\rm max}$, we have a degenerate case where $r_+$ and $r_{++}$ merge at $h_{\rm D_{+}}$ in Eq.~\eqref{bounds2}. Notice that $M^2_{\rm max}=\frac{1}{9\Lambda}$ for $a=0$, in agreement with the results for the de Sitter-Schwarzschild black hole. The other degenerate case, i.e., $M=M_{\rm min}$, occurs when $r_+$ and $r_-$ merge at $h_{\rm D_{-}}$ in Eq.~\eqref{bounds2}, which represents an extremal black hole, with
\begin{equation}
	a_{\rm ext}\sim\,M+\frac{\Lambda}{6}M^3\ .
\end{equation}
We see that the condition $a = M$ no longer leads to extremal black holes and that 
\begin{equation}
	a_{\rm ext}<a^{\rm KdS}_{\rm ext}\equiv\,M+\frac{\Lambda}{3}M^3\ ,
\end{equation}
where $a^{\rm KdS}_{\rm ext}$ corresponds to the extremal case for the KdS solution.
\par
 Notice that the line element~\eqref{newkerrds} shows a region where the spacetime violates the causality condition, as it happens for the standard solution in Eq.~\eqref{kdsstandard}. This takes place when $g_{\phi\phi}>0$, which defines a sector where the Killing field $\partial_\phi$ becomes timelike. This occurs adjacent to the ring singularity, in the region $r<0$, and goes from $r = 0$ to a maximum or minimum extension, corresponding to a maximum or extreme black hole, respectively. 

Regarding rotational effects, we see that the angular speed $\Omega\equiv\,-\frac{g_{t\phi}}{g_{\phi\phi}}$ for the Lense-Thirring effect
\begin{equation}
\Omega=\frac{ar(2M+\frac{\Lambda}{3}r^3)}{\rho^2(r^2+a^2)+a^2r(2M+\frac{\Lambda}{3}r^3)\sin^{2} \theta\ }<\Omega_{\rm KdS}\ ,
\end{equation}
which is much simpler than $\Omega_{\rm KdS}$, the corresponding one for the KdS metric in Eq.~\eqref{kdsstandard}. In this respect, although it is true that every rotating solution produces a frame dragging, the warped curvature displayed in Eq.~\eqref{R} cannot be generated by the Lense-Thirring effect in the KdS solution~\eqref{kdsstandard}. Also note that, in general, the line element~\eqref{newkerrds} describes the spacetime of a uniform thermal bath (regardless of its nature) surrounding a black hole in equilibrium. In this sense, the solution is not limited to the existence of a cosmological constant. 
\par
If we compare the event horizons $r_+$ of both solutions in~\eqref{kdsstandard} and~\eqref{newkerrds}, for a given value of $\{M,\,a,\,\Lambda\}$, we find that $r_+<r^{\rm KdS}_+$. This indicates a larger screening effect due to $\Lambda$ in the new solution. This agrees with the shadow of both solutions in Fig.~\ref{fig2}.
\par
Finally, if the Minkowski vacuum is filled by a source whose energy-momentum tensor is traceless, then there will be no warped curvature as shown in the expression~\eqref{R}. This explains why the curvature remains $R=0$ in the axially symmetric electrovacuum, namely, when we departure from the Reissner-Nordstr\"{o}m solution to generate the Kerr-Newman solution~\cite{Contreras:2021yxe}. The above has a direct and quite important consequence regarding alternative theories, which we express in the form of conjecture: {\it if there is a gravitational sector not described by general relativity, and it is conformal invariant, then a rotating black hole will produce no warped curvature of spacetime surrounding it.}

\section{Conclusions}
\label{con}
Following the scheme developed in Ref.~\cite{Contreras:2021yxe}, we find a rotating version of the Schwarzschild de Sitter spacetime, which represents a new solution describing the exterior of a black hole with cosmological constant. We find that the new solution, displayed in Eq.~\eqref{newkerrds},
besides being simpler and therefore easier to analyze than the standard one in Eq.~\eqref{kdsstandard}, presents a phenomenon never described, such as the warped curvature shown in~\eqref{R}. This rotational effect on the curvature $R$ will appear as long as the energy-momentum tensor filling the axialsymmetric vacuum has a nonzero trace, as in fact it is for the case of a cosmological constant.
\par
The new solution was examined in detail, precisely identifying the bonds for $M(a,\Lambda)$ in~\eqref{bounds}, within which the existence of BHs is possible. We want to stress that this new scheme is particularly attractive to implement in theories beyond Einstein, where in general finding exact axialsymmetric BH solutions is a difficult task, and in most cases impossible.
\par
We conclude emphasizing a critical point of particular importance related to the uniqueness theorem. In this paper, like Carter's solution, the spacetime surrounding rotating BHs contains only the cosmological constant, that is, the right-hand side of Eq.~\eqref{efe} is given by $\Lambda\,g_{\mu\nu}$.
In Carter's solution, $\Lambda$ remains constant and uniform everywhere. In our case, the Minkowski spacetime is filled by $T_{\mu\nu}=\Lambda\,g_{\mu\nu}$ generating a de Sitter spacetime with uniform energy density $\epsilon=\Lambda$. However, as soon as we get near to a rotating BH, we observe rotational effects on the energy density $\Lambda$ and spacetime curvature $R$, as is explicitly shown by Eqs.~\eqref{R},~\eqref{energyax} and~\eqref{pressuresax}. Of course, by construction, it is not possible to see these effects in the KdS solution. Indeed, it is difficult for a second solution to exist satisfying a condition as rigid as that imposed on Carter's solution. However, if we relax this condition, we can probably find new features of spacetime near rotating BHs, such as warped curvature, that are certainly worth studying.

%
%

%
%
%
\bibliography{references.bib}
\bibliographystyle{apsrev4-1.bst}
%
%
\end{document}